\documentstyle[11pt,newpasp,twoside,epsf]{article}
\def\edcomment#1{\iffalse\marginpar{\raggedright\sl#1\/}\else\relax\fi}
\reversemarginpar

\begin{document}
\title{Fabry-Perot Survey of Emission-Line Galaxies}
 \author{Sylvain Veilleux}
\affil{Department of Astronomy, University of Maryland, College Park, MD 20742}

\begin{abstract}
The recent results from a deep Fabry-Perot survey of nearby active and
star-forming galaxies are presented.  Line-emitting material is
detected over two orders of magnitude in galactocentric radius, from
the 100-pc scale of the active or starbursting nucleus out to several
10s of kpc, sometimes well beyond the optical confines of the host
galaxies.  The excitation and dynamical properties of the nuclear gas
are studied to constrain the impact of galactic winds on the host
galaxies and their environment. The properties of the warm ionized
material on the outskirts of galaxies provide important clues for
understanding galaxy formation and evolution. A new technique to
search for starburst-driven wind galaxy candidates is discussed. The
next generation of Fabry-Perot instruments on large telescopes
promises to improve the sensitivity of emission-line galaxy surveys at
least tenfold.
\end{abstract}

\section{Introduction}

The Fabry-Perot interferometer has the distinct advantage of providing
detailed spectrophotometric information over a larger field of view
(FOV) than that of other 3D instruments. The Fabry-Perot
interferometer is therefore ideally suited to study nearby galaxies
where the line-emitting gas extends over several arcminutes. This
ionized material is an excellent probe of the phenomena taking place
in the core of starburst and active galaxies, and can be used to
quantify the impact of nuclear and star-formation activity on the
environment and vice-versa.

Over the last decade, our group has used the scanning mode of the
Hawaii Imaging Fabry-Perot Interferometer (HIFI) on the CFHT 3.6m and
UH 2.2m and the TAURUS-II system on the AAT 3.9m and WHT 4.2m to study
in detail a sample of about twenty nearby starbursts and active
galaxies. A summary of the results from this portion of the survey is
presented in \S 2.  More recently, we have used a low-order
Fabry-Perot interferometer to obtain very deep emission-line maps of
several normal and active galaxies. This tunable-filter mode is
particularly efficient to search for warm ionized gas on the outskirts
of galaxies. The first results from this work are discussed in \S 3.
The tunable filter has also proven to be an efficient tool to search
for starburst galaxies with large-scale galactic winds. This technique
was used recently to uncover a shock-excited wind in the starburst
galaxy NGC~1482. The data are described in \S 4 along with possible
applications of this technique to search for wind candidates at higher
redshifts. Future avenues of research with Fabry-Perot interferometers
are briefly discussed in \S 5.

\section{Superwind Survey with a scanning Fabry-Perot Interferometer}

The initial portion of our galaxy survey focussed on obtaining
high-quality data cubes of starburst and active galaxies with known
galactic-scale outflows. All of the objects in the sample are nearby
($z <$ 0.01) to provide a good spatial scale ($<$ 200 pc
arcsec$^{-1}$), and present line-emitting regions which extends $>
30\arcsec$ to fully exploit the large FOV advantage of the Fabry-Perot
interferometer.  The Fabry-Perot data were combined with radio and
X-ray data and HST images to track the energy flow through various gas
phases (neutral, warm, hot, and relativistic). The high level of
sophistication of recent hydrodynamical simulations (e.g., Strickland
\& Stevens 2000) has provided the theoretical basis to interpret our
data and to predict the evolution and eventual resting place (disk,
halo, or intergalactic medium) of the outflowing material.  So far,
high-quality data sets have been obtained and analyzed for about
twenty galaxies (see recent review by Veilleux et al. 2001). These
outflows typically show the following optical properties (see examples
in Figs. 1 and 2):

\begin{figure}[htb]
\vskip 2.75in
\caption{Starburst-driven wind in NGC 3079.  H$\alpha$ + [N~II] line
emission map of the nuclear superbubble obtained with HST WFPC2 by
Cecil et al. (2001a). The superbubble is made of four separate bundles
of ionized filaments, curving up to $\sim$ 0.6 kpc above the disk,
then dispersing as a spray of droplets. A Fabry-Perot data cube of the
same region provides detailed kinematic information on the
line-emitting gas (Veilleux et al. 1994). A remarkably good agreement
is found between these data and the predictions of numerical
simulations, including the presence of material with large vortex
velocities near the top of the bubble indicative of a partially
ruptured structure. }
\end{figure}

\clearpage

\begin{itemize}

\item[$\bullet$] The optical winds in starburst galaxies are always
nearly perpendicular to the disk of the host galaxy (e.g., M~82,
NGC~3079; Shopbell \& Bland-Hawthorn 1998; Veilleux et al. 1994; Cecil
et al. 2001a; Fig. 1), while AGN-driven optical winds are often
lopsided and sometimes tilted with respect to the polar axis of the
host galaxy (e.g., NGC~4388; Veilleux et al. 1999; Fig. 2).

\item[$\bullet$] The solid angle subtended by these winds, $\Omega_W/4\pi$ 
$\approx$ 0.1 -- 0.5.

\item[$\bullet$] The radial extent of the line-emitting material
involved in the outflow, $R_W$ = 1 -- 5 kpc.

\item[$\bullet$] The outflow velocity of the line-emitting material,
$V_W$ = 100 -- 1500 km s$^{-1}$ regardless of the escape velocity of
the host galaxy. Until recently the record holder was NGC~3079, where
outflow velocities in excess of 1500 km s$^{-1}$ are directly measured
(Veilleux et al. 1994). But the outflow velocities in NGC~1068 are now
known to be far larger than this value (Cecil et al. 2001b).

\item[$\bullet$] The dynamical time scale, $t_{\rm dyn}$ $\approx$
$R_W/V_W$ = $10^6 - 10^7$ years.

\item[$\bullet$] The ionized mass involved in the outflow, $M$ =
10$^5$ -- 10$^7$ M$_\odot$, a relatively small fraction of the total
ISM in the host galaxy.

\item[$\bullet$] The ionized mass outflow rate, d$M$/d$t$ $\approx$
$M$/$t_{\rm dyn}$ = 0.1 -- 1 $M_\odot$ yr$^{-1}$ (i.e. much larger
than the mass accretion rate in AGN, or roughly equal to within a
factor of 10 to the star formation rate in starburst galaxies).

\item[$\bullet$] The kinetic energy of the outflowing ionized
material, $E_{\rm kin}$ = 10$^{53}$ -- 10$^{55}$ ergs, taking into
account both the bulk and ``turbulent'' (spectrally unresolved)
motions. This mechanical energy is equivalent to that of $\sim$ 10$^2$
-- 10$^4$ Type II SNe.

\item[$\bullet$] The kinetic energy rate of the outflowing ionized
material, d$E_{\rm kin}$/d$t$ $\approx$ $E_{\rm kin}/t_{\rm dyn}$ =
few $\times 10^{39} - 10^{42}$ ergs s$^{-1}$.

\item[$\bullet$] Evidence for entrainment of (rotating) disk material
is seen in several objects (e.g., Circinus, NGC~2992; Veilleux \&
Bland-Hawthorn 1997; Veilleux, Shopbell, \& Miller 2001).

\item[$\bullet$] The source of ionization of the line-emitting
material taking part in these outflows is diverse.  Pure
photoionization by the AGN can explain the emission-line ratios in
NGC~3516 (Veilleux, Tully, \& Bland-Hawthorn 1995) and NGC~4388
(Veilleux et al. 1999). Shock ionization is probably contributing in
Circinus and NGC~2992. Shock ionization often appears to be the
dominant process in starburst-driven winds (see \S 4).

\item[$\bullet$] Galactic winds in AGNs are roughly aligned with
galaxy-scale (several kpc) radio emission, sometimes encompassing what
appears to be poorly collimated radio ``jets'' (e.g., NGC~4388;
Fig. 2; Veilleux et al. 1999). These mass-loaded ``jets'' are the
probable driving engine for these winds. In most cases, however,
torus-driven winds cannot formally be ruled out (Veilleux et
al. 2001).  

\end{itemize}

\begin{figure}[htb]
\vskip 6in
\caption{AGN-driven wind in NGC 4388. Barycentric velocity field of
the line-emitting gas.  (top) H$\alpha$; (bottom) [O~III]
$\lambda$5007. North is at the top and east to the left. The optical
continuum nucleus is indicated in each panel by a cross. The velocites
range from about --200 km s$^{-1}$ to +200 km s$^{-1}$ relative to
systemic (= 2,525 km s$^{-1}$). The galactic disk is oriented in the
east-west direction and shows bar streaming near the center in
addition to circular rotation in the outer portions of the disk.  The
material above the disk does not follow normal galactic rotation; it
is outflowing from the nucleus. A redshifted cloud located 2 kpc to
the southwest of the nucleus is also taking part in the outflow.
Loosely collimated radio ``jets'' are at the origin of this bipolar
outflow (see Veilleux et al. 1999 for more detail). }
\end{figure}

\clearpage

\begin{itemize}
\item[$\bullet$] The excellent agreement between the predictions of
numerical simulations and our data, both in terms of morphology and
kinematics, confirms that the pressure gradient in the gaseous disk of
the host galaxy is the primary focussing mechanism in starburst-driven
winds.

\item[$\bullet$] Some of these galactic winds appear to have a strong
impact on the thermal and chemical evolution of the host galaxy and
even possibly beyond (e.g., NGC~3079; Veilleux et al. 1994; Cecil et
al. 2001a).
\end{itemize}

\section{New Tunable-Filter Survey of Active and Star-Forming Galaxies}

The optical data obtained during the early stage of the survey were
obtained with conventional high-order Fabry-Perot interferometers used
in a scanning mode, where the etalon is scanned over a range of
spacings to build up a spectral data cube (x, y, $\lambda$) over a
narrow spectral interval ($\sim$ 25 -- 50 \AA).  This technique
provides complete spatial and kinematic sampling of the superwinds, an
important condition to reconstruct their tridimensional
structure. However, an important limitation of this technique is the
relatively small effective FOV of the Fabry-Perot data, which is
determined primarily by the range in gap spacings covered by our data
cubes rather than the FOV of the instrument. Our data cubes typically
contain only 30 -- 50 images (limited by the time spent on each
object) and therefore cover only about $\pm$ 1$\arcmin$ ($\pm$ $\sim$
3 kpc) above and below the galaxy plane of our targets.  Emission
beyond this region falls outside the velocity range of our Fabry-Perot
data and remains undetected.

Low-order etalons such as those used in the Taurus Tunable Filter (TTF; 
see Bland-Hawthorn \& Jones 1998) can remedy the
situation, providing gains of about an order of magnitude in both the
effective FOV ($\sim$ 10$\arcmin$ in {\em one} exposure) and H$\alpha$
limiting surface brightness ($\sim$ a few x 10$^{-18}$ erg s$^{-1}$
cm$^{-2}$ arcsec$^{-2}$ {\em i.e.} about an order of magnitude fainter
than that of published data on emission-line galaxies).  For the past
two years, our group has used the TTF on the AAT and WHT to map
several active and starburst galaxies as well as a number of normal
(``quiescent'') edge-on disk systems. The portion of the survey on
normal galaxies is part of a Ph.D. thesis by University of Maryland
student Scott T. Miller.  The results so far can be summarized as
follows (see review by Veilleux 2001; see also Fig. 3):
\begin{itemize}

\item[$\bullet$] In active and starburst galaxies, ionized filaments
often extend out to several tens of kpc, sometimes beyond the H~I edge
of the host galaxy.

\item[$\bullet$] Filamentary complexes are seen extending a few kpc
above and below the disks of normal star-forming galaxies.  Both the
mass and extent of the extraplanar material in these galaxies appear
to be correlated with the {\em local} surface density of star
formation activity in the disk.

\item[$\bullet$] Emission-line ratio maps constructed from multi-line
imaging of these objects reveal line ratios which are not H~II
region-like. An early analysis of our results on normal disk galaxies
combined with complementary long-slit spectra suggests that the
extraplanar gas is primarily photoionized by the highly diluted and
filtered radiation from OB stars in the disk (Miller 2002). However,
an additional source of ionization is often needed to reproduce all of
the vertical line ratio profiles. In active galaxies, the primary
source of ionization of the extended nebula is either the AGN itself
or shock excitation.

\item[$\bullet$] Multi-line imaging slightly shifted in velocity space
provides strong constraints on the kinematics of the warm ionized gas.
In all cases studied so far, the gas appears to be bound to the host
galaxy.
\end{itemize}

\section{Search for Starburst-Driven Winds with the Tunable Filter}

\subsection{Discovery of a Galactic Wind in the Starburst Galaxy NGC~1482}

In the course of our TTF survey of emission-line galaxies, we
discovered a remarkable emission-line structure in the early-type
spiral galaxy NGC~1482 (Veilleux \& Rupke 2002). The TTF data are
shown in Figure 3$d$ and in more detail in Figure 4. Strong H$\alpha$
and [N~II] emission is detected along the plane of the host galaxy
(P.A. $\approx$ 103$^\circ$). In addition, an hourglass-shaped
structure is seen in both H$\alpha$ and [N~II] $\lambda$6583,
extending along the minor axis of the galaxy at least $\sim$ 1.5 kpc
above and below the galactic plane. This structure is more easily
visible in [N~II] $\lambda$6583 than in H$\alpha$. This is
particularly apparent in the lower left panel of Figure 4, where we
present a [N~II]/H$\alpha$ ratio map of this object. The [N~II]
$\lambda$6583/H$\alpha$ ratios measured in the disk of the galaxy
[$\approx$ 0.3 (outer disk) -- 0.6 (inner disk)] are typical of
photoionization by stars in H~II regions, but the ratios in the
hourglass structure are 3 -- 7 times larger ([N~II]
$\lambda$6583/H$\alpha$ $\approx$ 1.0 -- 2.3). This ratio of a
collisionally excited line to a recombination line is fundamentally a
measure of the relative importance of heating and ionization (e.g.,
Osterbrock 1989).  The most likely explanation for these unusual line
ratios is shock ionization resulting from the interaction of an
energetic large-scale outflow with the ambient material of the host
galaxy.

Complementary long-slit spectra obtained with the dual-beam
spectrograph (DBS) on the MSSSO 2.3m telescope confirm that the
hourglass structure is due to a large-scale galactic wind and that the
extreme line ratios are produced through shocks (Fig. 5). Line
splitting of up to $\sim$ 250 km s$^{-1}$ is detected along the axis
of the hourglass structure out to at least 16\arcsec\ (1.5 kpc) above
and below the galaxy disk. Normal galactic rotation dominates the
kinematics of the gas within 5 -- 6\arcsec\ ($\sim$ 500 pc) from the
disk. Maximum line splitting often coincides with regions of low
emission-line surface brightness. These results can be explained if
the extraplanar emission-line material forms a biconical
edge-brightened structure which is undergoing outward motion away from
the central disk. In this case, the blueshifted (redshifted)
emission-line component corresponds to the front (back) surface of the
bicone. The lack of obvious velocity gradient in the centroid of the
line emission suggests that the main axis of the bicone lies close to
the plane of the sky. The fact that the amplitude of the line
splitting does not decrease significantly with distance from the
galaxy indicates that the entrained material is not experiencing
significant deceleration (i.e., it is a blown-out wind).  A lower
limit of 2 $\times$ 10$^{53}$ ergs is derived for the kinetic energy
of the outflow based on the deprojected gas kinematics and the amount
of ionized material entrained in the outflow. We find that the
starburst at the

\begin{figure}[htb]
\vskip 4.0in
\caption{Tunable filter images of nearby galaxies. North is up, east
to the left. {\bf (a)} One-hour H$\alpha$ image of the archetypical
Seyfert 2 galaxy NGC~1068. Line emission is detected for the first
time out to $\sim$ 12 kpc from the nucleus, slightly beyond the H~I
edge of this galaxy. The biconical geometry of the complex and the
relative strengths of the emission lines suggest that the central AGN
is contributing to the ionization of this material.  (Shopbell,
Bland-Hawthorn, \& Veilleux 2002, in prep.); {\bf (b)} A line-emitting
filament is detected at 19 kpc from the nucleus of NGC 7213.  This
filament was independently discovered by Hameed et al. (2001), and
seems to coincide spatially and kinematically with tidal debris seen
in H~I 21 cm. (Veilleux \& Rupke 2002, in prep.); {\bf (c)} H$\alpha$
of the field surrounding the quasar MR~2251--178. A spiral-like nebula
extending $\sim$ 200 kpc around the quasar is detected in this
image. This nebula is ionized by the quasar and in rotation around it
(see Shopbell, Veilleux, \& Bland-Hawthorn 1999 for more detail).
{\bf (d)} [N~II] $\lambda$6583/H$\alpha$ ratio in the starburst galaxy
NGC~1482. The [N~II]/H$\alpha$ ratio is $<$ 1 in the galactic disk (PA
= 100$^\circ$) but $>$ 1 in the material above the disk. These data
suggest the presence of a shock-excited superwind in NGC~1482. See \S
4 and Veilleux \& Rupke (2002) for more detail.  }
\end{figure}

\begin{figure*}[ht]
\vskip 6.0in
\caption{ Narrow-band images of NGC~1482 obtained with the TTF:
(clockwise from upper left corner) red continuum, [N~II] $\lambda$6583
line emission, H$\alpha$ line emission, and [N~II]
$\lambda$6583/H$\alpha$ ratio.  The panels on the right are on a
logarithmic intensity scale, while those on the left are on a linear
scale.  North is at the top and east to the left.  The positions of
the continuum peaks are indicated in each image by two crosses. The
spatial scale, indicated by a horizontal bar at the bottom of the red
continuum image, is the same for each image and corresponds to $\sim$
21\arcsec, or 2 kpc for the adopted distance of 19.6 Mpc for NGC~1482.
The [N~II] $\lambda$6583/H$\alpha$ ratio is below unity in the galaxy
disk but larger than unity in the hourglass-shaped nebula above and
below the disk. This structure is highly suggestive of a galactic
wind.}
\end{figure*}

\clearpage

\begin{figure*}[ht]
\vskip 6.5in
\caption{ Sky-subtracted long-slit spectra obtained parallel to the
galactic disk (P.A. $\sim$ 103$^\circ$). For each panel, south-east is
to the left and north-west to the right. The spectra displayed on the
left are offset to the north-east by 0\arcsec\ (bottom panel),
9\arcsec\ -- 10\arcsec, 11\arcsec\ -- 12\arcsec, and 13 -- 14\arcsec\
(top panel) from the major axis of the host galaxy. The spectra
displayed on the right are offset by approximately the same quantity
in the south-west direction. The vertical segment in each panel
represents 500 km s$^{-1}$. The presence of line splitting above and
below the disk confirms the presence of a large-scale wind in this
galaxy. }
\end{figure*}

\clearpage

\noindent
base of the wind nebula is powerful enough to drive the outflow, based
on the star formation rate derived from the $IRAS$ database. A more
detailed discussion of the TTF data may be found in Veilleux \& Rupke
(2002).

\subsection{On the Use of Excitation Maps to Identify Starburst-Driven Wind
Galaxies}

The traditional method of identifying galaxy-scale winds in starburst
galaxies is to look for the kinematic signature (e.g., line splitting)
of the wind along the minor axis of the host galaxy disk. Edge-on disk
orientation reduces contamination of the wind emission by the
underlying disk material and facilitates the identification. This
method is time-consuming since it requires deep spectroscopy of each
candidate wind galaxy with spectral resolution of $\la$ 100 km
s$^{-1}$. Line ratio maps like the ones shown in Figures 3$d$ and 4
represent a promising new way to detect galactic winds in starburst
galaxies. The line ratio method only requires taking narrow-band
images of candidate wind galaxies centered on two (or more) key
diagnostic emission lines which emphasize the contrast in the
excitation properties between the shocked wind material and the
star-forming disk of the host galaxy.  [N~II] $\lambda$6583/H$\alpha$,
[S~II] $\lambda\lambda$6716, 6731/H$\alpha$, and [O~I]
$\lambda$6300/H$\alpha$ are the optical line ratios of choice for $z
\la 0.5$ galaxies (these ratios are enhanced in the wind of NGC~1482),
while [O~II] $\lambda$3727/H$\beta$ and [O~II] $\lambda$3727/[O~III]
$\lambda$5007 could be used for objects at larger redshifts.  The
spatial resolution of these images must be sufficient to distinguish
the galaxy disk from the wind material. Using NGC~1482 as a template,
we find that high-[N~II]/H$\alpha$ winds in edge-on starburst galaxies
would still be easily detected out to a distance of $\sim$ 200 Mpc
under 1\arcsec\ resolution. Imagers equipped with adaptive optics
systems should be able to extend the range of these searches by an
order of magnitude.

This method relies on the dominance of shock excitation in the optical
line-emitting wind component.  Surveys of local powerful wind galaxies
(e.g., Heckman, Armus, \& Miley 1990; Bland-Hawthorn 1995; Veilleux et
al. 1995; Lehnert \& Heckman 1996; Veilleux 2001) confirm that shocks
generally are the dominant source of excitation in the wind
material. These shock-dominated wind nebulae present line ratios which
are markedly different from the star-forming disks of the host
galaxies.  The superbubble in NGC~3079 presents large [N~II]
$\lambda$6583/H$\alpha$ ratios, reaching values of $\sim$ 3 at the
base of the bubble where the widths of the emission lines exceed 400
km s$^{-1}$. This is another clear case of shock-excited wind
triggered by a nuclear starburst. At the other end of the excitation
spectrum is the wind in M~82. The [N~II]/H$\alpha$ map of the southern
wind lobe of M~82 (Fig. 4 of Shopbell \& Bland-Hawthorn 1998) presents
two distinct fan-like structures with H~II region-like ratios
originating from the two bright star-forming regions in the disk of
this galaxy. The line ratio technique would not be able to distinguish
between line emission from this type of photoionization-dominated wind
nebulae and contamination from a star-forming disk seen nearly
face-on.  Blind searches for galactic winds based on the excitation
contrast between the disk and wind components would therefore favor
the detection of winds in edge-on hosts where the wind component is
not projected onto the disk component. This orientation bias would
need to be taken into account to get a complete census of
starburst-driven wind galaxies.  AGN-driven winds may also contaminate
samples selected from excitation maps if the spatial resolution is not
sufficient to separate the active nucleus from the disk material.

\section{Future}

The contribution of J. Bland-Hawthorn to these proceedings lists large
telescopes with planned or proposed Fabry-Perot interferometers.
Tunable filters on 6 -- 10m telescopes will be ideally suited to
survey the line-emitting universe at low and high redshifts.  This is
particularly true for the Maryland-Magellan Tunable Filter
(MMTF) proposed for the Magellan 6.5m telescope which will combine wide field
of view (monochromatic spot with diameter $\sim$ 10$\arcmin$ --
27$\arcmin$) with outstanding narrow-band imaging capabilities over a
broad range in wavelengths (5000 -- 9200~\AA) and bandpasses (10 --
100~\AA). The exceptional areal coverage and sensitivity of this
instrument will result in a gain of at least an order of magnitude in
survey efficiency. MMTF will be a very efficient tool to study star
formation over a broad range of redshifts and galaxy environments.

\vskip 0.3in

\acknowledgements The results presented in this paper are part of a
long-term effort involving many collaborators, including
J. Bland-Hawthorn, G. Cecil, P. L. Shopbell, and R. B. Tully and
Maryland graduate students S. T. Miller and D. S. Rupke. This work is
partially supported by a Cottrell Scholarship awarded by the Research
Corporation, NASA/LTSA grant NAG 56547, and NSF/CAREER grant
AST-9874973.

\end{document}